\documentclass[onecolumn,preprint,12pt,3p,authoryear]{aastex631}

\definecolor{mygray}{gray}{0.9}
\usepackage{graphicx}
\usepackage[colorlinks=true]{hyperref}
\usepackage[utf8]{inputenc}
\usepackage{amsmath} %defines a lot of math symbols
\usepackage{xcolor}
\usepackage{threeparttable}
\usepackage{lipsum}
\usepackage{mathtools}
\usepackage{cuted}

\usepackage{booktabs}

\begin{document}

\title{Spatiotemporal Low FIP Abundance: A Catalyst for Coronal Condensation}

% at Mānoa
\author[0009-0006-9652-0036]{Luke Fushimi Benavitz}
\affil{Institute for Astronomy, University of Hawai`i, Honolulu, HI 96822, USA}
\author[0000-0003-4739-1152]{Jeffrey W. Reep}
\affil{Institute for Astronomy, University of Hawai’i, Pukalani, HI 96768, USA}
\author[0000-0002-8259-8303]{Lucas A. Tarr}
\affil{National Solar Observatory, 22 Ohi‘a Ku St., Makawao, HI 96768, USA}
\author[0000-0003-0774-9084]{Andy S.H. To}
\affil{ESTEC, European Space Agency, Keplerlaan 1, PO Box 299, NL-2200 AG Noordwijk, The Netherlands}

\begin{abstract}

Radiative losses play a critical role in the cooling of plasmas. When chromospheric plasma is sufficiently heated, it can flow into coronal loops which subsequently cool down due to radiation. From observations, we infer that this cooling does not occur uniformly, often resulting in coronal condensations such as coronal rain. To date, coronal condensations have only been found in simulations of steadily-heated loops, and never in impulsively-heated ones. We implement spatiotemporally variable elemental abundances in a radiative hydrodynamic code. Flows, including chromospheric evaporation, directly cause a shift in the local elemental abundances, which then affects the local radiative loss rate. As a consequence, we find that incorporating spatiotemporal low FIP elemental abundances into coronal loop simulations directly causes coronal condensations, which are otherwise absent in impulsively heated loop or flare models. We conclude that spatiotemporal variations in elemental abundances are a fundamental feature of the solar corona and are therefore necessary to accurately model radiation.

\end{abstract}


\section{Introduction}
\label{s.Introduction}

Radiation is a critical process in the cooling of plasmas, including the solar corona. %critical point
Observations of the corona have shown that elemental abundances vary in space and time \citep{Brooks2015,DelZanna2021,To2021,Mondal2021,Nama2023,Mondal2023,Suarez2023,Brooks2024}, but models typically assume that they are fixed in space and time (\textit{e.g.}, \citealt{Reep2020}). %motivation
Since radiative energy loss is directly influenced by these abundances, this assumption oversimplifies the physics governing the solar atmosphere in our models \citep{Reep2024}. 

Bolometric radiative losses are also crucial to understanding how both heated loops and flares will evolve. As we do not generally know the local abundances (\textit{e.g.}, $n_{Fe} / n_{H}$) at a given position along the loop, it is not understood exactly how radiation varies with time and space. We rely on our understanding of radiation to infer important quantities such as emission, spectral line intensities, densities, temperatures, and velocities in loops and flares. For optically thin plasmas, radiative losses are described by

\begin{equation}
    R = - n_e n_H \Lambda(T_e , n_e, f)
    \label{eq.BolLoss}
\end{equation}

\noindent where $n_e$ is number density of electrons, $n_H$ is the number density of hydrogen, and $\Lambda(T_e , n_e, f)$ is the total emissivity of the plasma \citep{RTV1978,Cook1989,Mason1994,Bradshaw2013Rad}. $\Lambda$ is a function of temperature, density, and elemental abundances. We can describe local elemental abundances via the abundance factor \textit{f}

\begin{equation}
    f = \frac{n_{X} / n_{H}} {n_{X,phot} / n_{H,phot}}
    \label{eq.abdFactor}
\end{equation}

\noindent which is a measure of how enhanced or diminished a given element X is relative to hydrogen, and at some location relative to its average photospheric value.  

In the corona, \textit{f} changes due to flows of a given element.  We expect changes in coronal abundances when there is a sufficiently strong heating event that causes those flows. Energy is transferred from the corona down to the upper chromosphere which causes local pressures in the chromosphere to rise, and this results in flows of chromospheric material along magnetic field lines through coronal loops into the corona called chromospheric ``evaporation" \citep{Hirayama1974,Acton1982,Fisher1984,Fisher1985,Antonucci1984,Antonucci1985}. In this work, we will use $f$ to describe all elements with low first ionization potential (FIP) as a collective, rather than treating individual elements. 

There is an observed enhancement of elements with low FIP of $\leq 10$ eV (\textit{e.g.}, Fe, Si, Mg) in the solar corona compared to their photospheric abundances, while elements with high FIP of $> 10$ eV (\textit{e.g.}, He, O, Ar) remain largely unchanged \citep{McKenzie1992,Feldman1992}, which is called the FIP effect. Emission lines from both low and high FIP elements provide diagnostics of coronal abundances \citep{Pottasch1964,DelZanna2014}.
The FIP effect can be measured using high-resolution spectroscopic observations in extreme ultraviolet (EUV) and X-ray wavelengths \citep{Brooks2022,Mondal2023}, from instruments such as Hinode's EUV Imaging Spectrometer (Hinode/EIS; \citealt{culhane2007}), as done for example in \citet{Baker2013} and \citet{Brooks2015}, and Solar Orbiter’s Spectral Imaging of the Coronal Environment (SPICE; \citealt{spice2020}), as in, \textit{e.g.},  \citet{Brooks2024SPICE}.

Observations show that coronal loops emerge at photospheric abundances, and, after time, fractionate, \textit{i.e.}, low FIP elements become enhanced in the corona relative to photospheric abundances\citep{Widing2001,Baker2018,Lodders2025}. 
The leading explanation for the FIP effect involves the ponderomotive force generated by Alfvén waves propagating from the corona
down through the transition region and refracting within the chromosphere, which preferentially accelerates ions into the corona while leaving neutrals largely unaffected \citep{Laming2015, Laming2021}. 
In the chromosphere, almost all low-FIP elements are ionized and high-FIP elements are largely neutral. \citet{Brooks2018} showed that, depending on where low or high FIP elements are depleted or enhanced, there is a significant difference in the cooling timescale of the loop. 
Understanding the FIP effect provides insight into both plasma fractionation processes and the composition of the solar wind, which in turn influences space weather and heliospheric dynamics. For clarity's sake, we do not directly model the FIP effect in this work, only the variation in abundance. 

Spatial variation in the elemental abundances causes a corresponding variation in the radiative loss rate.  If the gradient in the loss rate is large enough, it can cause what is called a coronal ``condensation" event \citep{Muller2003,Peter2012,Li2018,MartinezGomez2020,Li2021}. For example, coronal rain can occur where clumps of relatively cool and dense plasma fall back down to the chromosphere.
Rain occurs when there is a localized spike in radiation, \textit{i.e.} when there is a higher cooling rate at one location, causing the temperature to plummet rapidly and the radiative loss rate to grow even stronger.  This thermal runaway could be triggered by a local spike in density%($R \propto n^2 \Lambda$)
, a local imbalance in heating and cooling%($R > H$)
, a local dip in temperature, or, as we will show, a local peak in the abundance of certain elements \citep{Hildner1974,Klimchuk2019peak,Antolin2020}. 

Several specific mechanisms can create these conditions. The most well studied process is Thermal Non-Equilibrium (TNE) \citep{Kuin1982,Antiochos2000,Antolin2010,Antolin2015,Froment2015,Froment2018,Johnston2019,Scott2024}, where if heating of the footpoints of a coronal loop persists on the order of an hour or more, a density and temperature gradient steep enough to cause coronal rain occurs \citep{Klimchuk2019}. This was supported through observations of long period pulsations with time scales of days reported by \citet{Auchere2018} and \citet{Froment2017,Froment2020}. 
Rain can also occur through a combination of TNE and impulsive heating \citep{Reep2020}. \citet{Li2022} and \citet{Jercic2023} showed that magnetic cross-field interaction can possibly cause rain. 
Recently, \citet{Kucera2024} found that repeated nanoflares at alternating footpoints could cause rain, and \citet{Yoshihisa2024} used shock waves to trigger the thermal runaway causing rain. 

\citet{Klimchuk2019peak} discussed the difference between TNE and thermal instability, namely that TNE refers to a situation where a loop fundamentally cannot find an equilibrium between heating and radiation, whereas a thermal instability refers to a situation where there exists an equilibrium that can be readily perturbed away \citep{Field1965}.  In both cases, a thermal runaway can be triggered that results in coronal condensations such as coronal rain (see also \citealt{Antolin2020}).

Although we observe rain in almost all solar flares \citep {Foukal1978, Jing2016,Mason2022}, models of flares have not found coronal rain with standard assumptions in simulations \citep{Reep2020}. 
Standard flare models assume energy deposition by electron beams, which causes strong chromospheric evaporation and fills the corona with hot plasma \citep{Brown1971,Emslie1978,Nagai1984}. The loop then cools primarily through conduction initially, which efficiently redistributes heat and radiation. As the loop isotropizes and the temperature drops, radiation becomes the dominant cooling mechanism and flows start to carry mass and energy out of the loop (enthalpy flux cooling; \citealt{bradshaw2010}).
Conduction smooths out any density or temperature gradients. This results in little spatial variation in radiation in the corona that prevents the thermal runaway needed for rain to form, so condensations are typically absent from impulsively heated simulations. 

The lack of rain could result from an oversimplification in the physics of loop models, with the use of spatially homogeneous and temporally static abundances in current models being one clear problem. 
Recently, \citet{Reep2024} implemented a time-variable abundance factor into the 0D ebtel++ code \citep{Klimchuk2008,Cargill2012a,Cargill2012b,Barnes2016} that models the changes due to heating events, causing chromospheric evaporation, in order to understand how this affects coronal loop cooling. 
They found that if the initial heating rate was weaker than $\sim 1 $ erg/s/cm$^3$, the variable abundance factor would cause the loop cooling time to change significantly. 

Inspired by \citet{Reep2024}, we have implemented variable elemental abundances into a higher dimensional model than the 0D ebtel++ code. We use the open-source HYDrodynamics and RADiation (HYDRAD) Code \citep{Bradshaw2003, Bradshaw2013}. HYDRAD is a field-aligned model that examines the hydrodynamics of a two-fluid plasma flowing along a magnetic flux tube, and it implements the full treatment of thermodynamics which includes optically thick chromospheric radiation \citep{Carlsson2012}, thermal conduction with flux limiting terms, and optically thin radiative losses in the corona. We calculate optically thin radiative losses with version 10 of the CHIANTI atomic database \citep{Dere1997,DelZanna2021}. In this paper, we implement a variable abundance factor $f(t, s)$ that modifies the abundance of low-FIP elements, and thus generalizes the calculation of radiation. Our method is analogous to the non-equilibrium ionization calculation implemented in \citet{Bradshaw2003} where they solve a continuity equation to calculate ion fractions. However, we solve for elemental abundances rather than ion fractions. This model will help us better probe the physics of loop and flare cooling.  

The layout of the paper is as follows: Section~\ref{s.Time Variable Low FIP Abundances} shows the implementation of spatial and time-variable abundance factor, Section~\ref{s.Loop Simulations} presents the simulations, the results are in Section~\ref{s.Results}, and implications are discussed in Section~\ref{s.Discussion and Conclusion}.

\section{Time Variable Low FIP Abundances}
\label{s.Time Variable Low FIP Abundances}

We modify the radiative loss rate in HYDRAD by incorporating a spatial and time-variable abundance factor $f(t, s)$, where $f(t,s)$ for all low-FIP elements are assumed to vary together. To find how $f$ changes with space and time, we can employ a continuity equation
\begin{equation}
    \frac{\partial{(\rho f)}}{\partial{t}} + \frac{1}{A} \frac{\partial}{\partial{s}}(\rho f A v) = 0
    \label{eq1}
\end{equation}
\noindent where $f$ is abundance factor, $\rho$ is plasma mass density, $A$ is cross-sectional area of the loop, $v$ is the bulk flow velocity, $t$ is time, and $s$ is the spatial coordinate along the loop, where $s=0$ is the photosphere. Equations of this form are sometimes referred to as color equations \citep{LeVeque_2002}, often used in the context of fluid mixing (\textit{e.g.}, \citealt{Xu1995}). We can rewrite Equation~\ref{eq1} using the chain rule
\begin{equation}
    f \frac{\partial{\rho}}{\partial{t}} + \rho \frac{\partial{f}}{\partial{t}} = - \left( \frac{f}{A} \frac{\partial{(\rho A v)}}{\partial{s}} + \rho v \frac{\partial{f}}{\partial{s}} \right).
    \label{eq2}
\end{equation}
\noindent  We can use the continuity equation for mass density,
\begin{equation}
    \frac{\partial{\rho}}{\partial{t}} + \frac{1}{A} \frac{\partial}{\partial{s}}(\rho A v) = 0,
    \label{eq3}
\end{equation}
\noindent and, by combining Equation~\ref{eq2} and Equation~\ref{eq3}, we find
\begin{equation}
    - \frac{f}{A} \frac{\partial{(\rho A v)}}{\partial{s}} + \rho \frac{\partial{f}}{\partial{t}} = - \frac{f}{A} \frac{\partial{(\rho A v)}}{\partial{s}} - \rho v \frac{\partial{f}}{\partial{s}}.
    \label{eq4}
\end{equation}
\noindent Simplifying further, we find that $f$ evolves in space and time according to the advection equation

\begin{equation}
    %\partial_{t} f = - v\ \partial_{s} f
    \frac{\partial f}{\partial t} + v \frac{\partial f}{\partial s} = 0.
    \label{eqf}
\end{equation}

Inspecting Equation~\ref{eqf}, $f$ only changes in time due to flows if there is a spatial gradient (i.e, $f(s) \neq$ const.). 
We solve this equation numerically with a forward-time center-space (FTCS) solver, using a staggered leapfrog integration, following the standard calculation in HYDRAD (see the appendix of \citealt{Bradshaw2013} for details). 
This will not simulate the FIP effect as there is no source term (\textit{e.g.}, a ponderomotive force), and therefore the low FIP elements flow at the bulk flow speed. 

We define that where $f = 1$, the elemental abundances are photospheric \citep{Asplund2009}, and that, elsewhere, elements with FIP below a threshold of 10 eV are enhanced at a given position $s$ by $f(t, s)$. The solution of Equation~\ref{eqf} is used to calculate $\Lambda(T_{e}, n_{e}, f)$ in Equation~\ref{eq.BolLoss} at each time step and location along the loop. The radiative losses are then calculated as done normally in HYDRAD, \textit{i.e.}, by calculating the emissivity in each grid cell with the CHIANTI atomic database \citep{Dere1997,DelZanna2021}. 

\section{Loop Simulations}
\label{s.Loop Simulations}

We tested the effect of including variable elemental abundances under two common loop heating scenarios: impulsive nanoflare heating (typical active region loop simulations) and electron beam heating (typical flare loop simulations). For both sets of simulations, 
the simulation time is 1 hour for a semi-circular loop length of $s=2L=50$ Mm, and the transition region is set at $s=2.26$ Mm. We use a VAL-C chromosphere model, which also serves as the boundary condition for the coronal portion of the loop. The initial conditions in the corona are stitched to the chromosphere by solving the hydrostatic equilibrium equations across the coronal portion of the loop, resulting in a coronal temperature of $\sim5\cdot10^5$ K and coronal density of $\sim10^8$ cm$^3$ \citep{Vernazza1981}. The impulsive nanoflare heating starts at the beginning of the simulation with a step function rise and decay of 200 s in total at a rate of $H_{TE} = 0.1$ erg/s/cm$^3$ uniformly over the whole loop. For the electron beam heated case, the beam starts at the beginning of the simulation and lasts 20 seconds with a cutoff energy of $E_{cut} = 15$  keV and constant flux of $F_{beam} = 2\cdot10^{10}$ erg/s/cm$^2$.

HYDRAD has an option for adaptive mesh refinement (AMR) which adjusts adaptive re-gridding to ensure adequate spatial resolution \citep{Bradshaw2013}. Any grid cell in the simulation can refine successively by splitting a cell into two (exactly half a cell width) when the gradients are above a defined threshold. Following similar criteria, the cells can merge together where spatial resolution is not needed. For example, \citet{Johnston2019} examined AMR in the context of TNE heating. For our simulations, we select an AMR level of 12 (\textit{i.e.}, any particular grid cell can split up to 12 times). 

We run these simulations for three different abundance cases: uniform and constant abundances using the coronal values given by \citet{Feldman1992}, uniform and constant abundances using the photospheric values given by \citet{Asplund2009}, and spatiotemporally variable abundances described by Equation~\ref{eqf} (where we use the values of \citealt{Asplund2009} for $f=1$). For the variable abundance case, we assume that the loop is initially fractionated, where $f = 1$ in the photosphere and $f = 4$ in the corona using a step function at the base of the transition region. Radiative losses are calculated using H, He, C, N, O, Ne, Na, Mg, Al, Si, S, Ar, Ca, Fe, and Ni, which are the 15 most abundant elements in the solar corona \citep{Asplund2009}.  We also use the formulation from \citet{Carlsson2012} for optically thick radiation in the chromosphere.  We assume no coronal background heating term ($H_B = 0$). 
%Radiation indirectly depends on many factors such as the initial heating event, cross sectional loop area, and elemental abundances, so we only vary the type of heating and elemental abundances.

\section{Results}
\label{s.Results}

\begin{figure*}%[h!]
    \centering
    \includegraphics[width=1\textwidth]{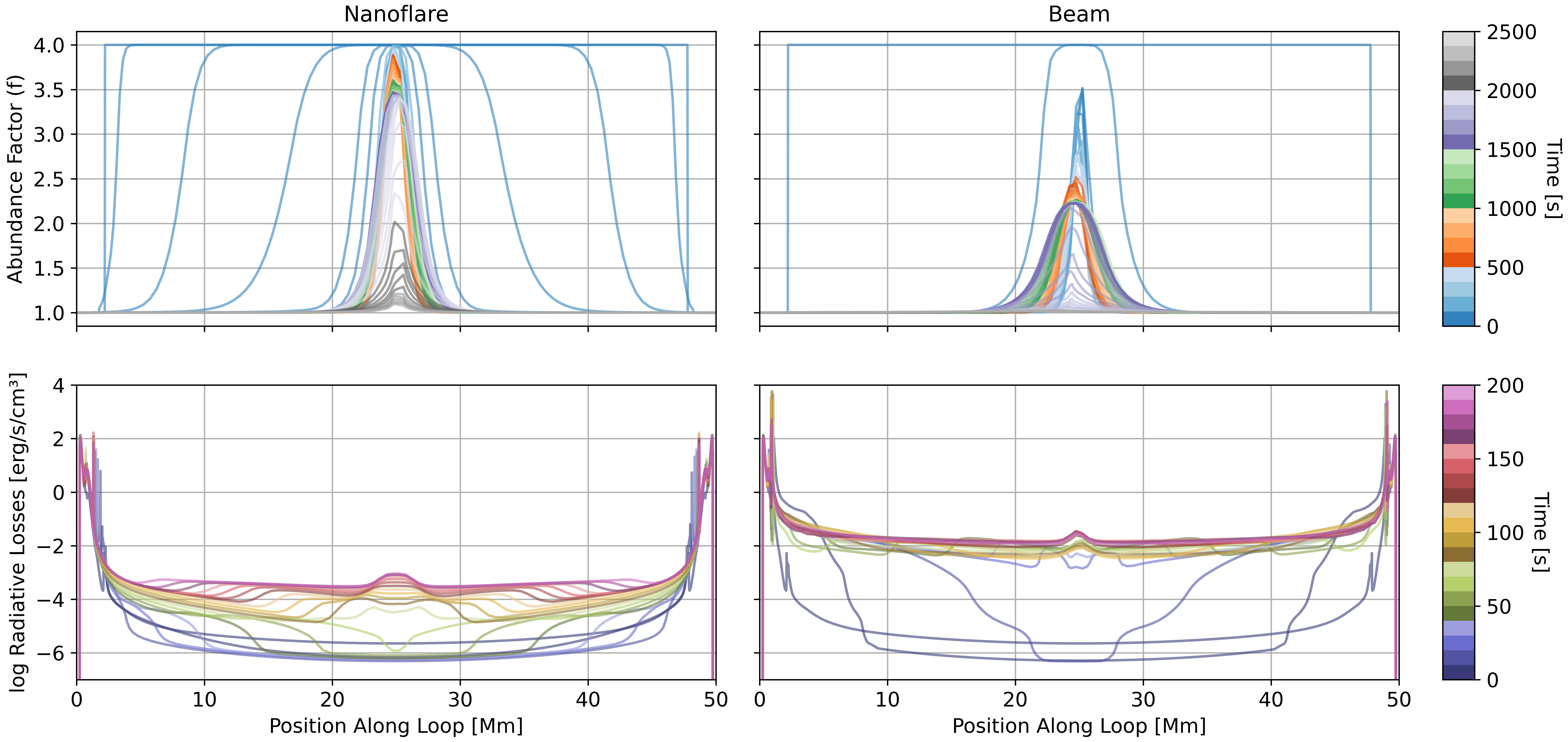}
    \caption{
    The top row shows the abundance factor as a function of position on the loop (for the full duration of the simulation), and the bottom row similarly shows the radiative losses (for the first 200 s). The left column shows the nanoflare-heated case, while the right is the beam-heated case.
    The corona starts at coronal abundances, but the abundance factor decreases rapidly in the corona due to chromospheric evaporation. The initial low FIP material is pushed to the apex, however, leaving a localized spike in $f$, and thus in radiation.
    }
    \label{fig.abdF}
\end{figure*}

Figure~\ref{fig.abdF} shows the simulations with variable elemental abundances. The upper row displays plots of abundance factor as a function of time and space, and the bottom row similarly shows the initial radiative losses up to $t=200$ seconds. The left column shows the nanoflare-heated case, while the right is the beam-heated case. Initially, the coronal part of the loop is fractionated, and, as time goes on, there are inflows that bring chromospheric material into the corona, thus reducing $f$. The two upward-streaming flows from each loop leg push low FIP material towards the apex, creating a local spike in $f$ that only decreases after outflows start to occur. This peak of \textit{f} at the apex of the loop only appears in the variable abundance cases, and results in a local peak in radiative losses at the apex of the loop in accordance with Equation~\ref{eq.BolLoss}.%, and the bottom row of plots show the radiation from the initial evaporation outflow.

\begin{figure*}[p]%[ht!]
    \centering
    \includegraphics[width=1\textwidth]{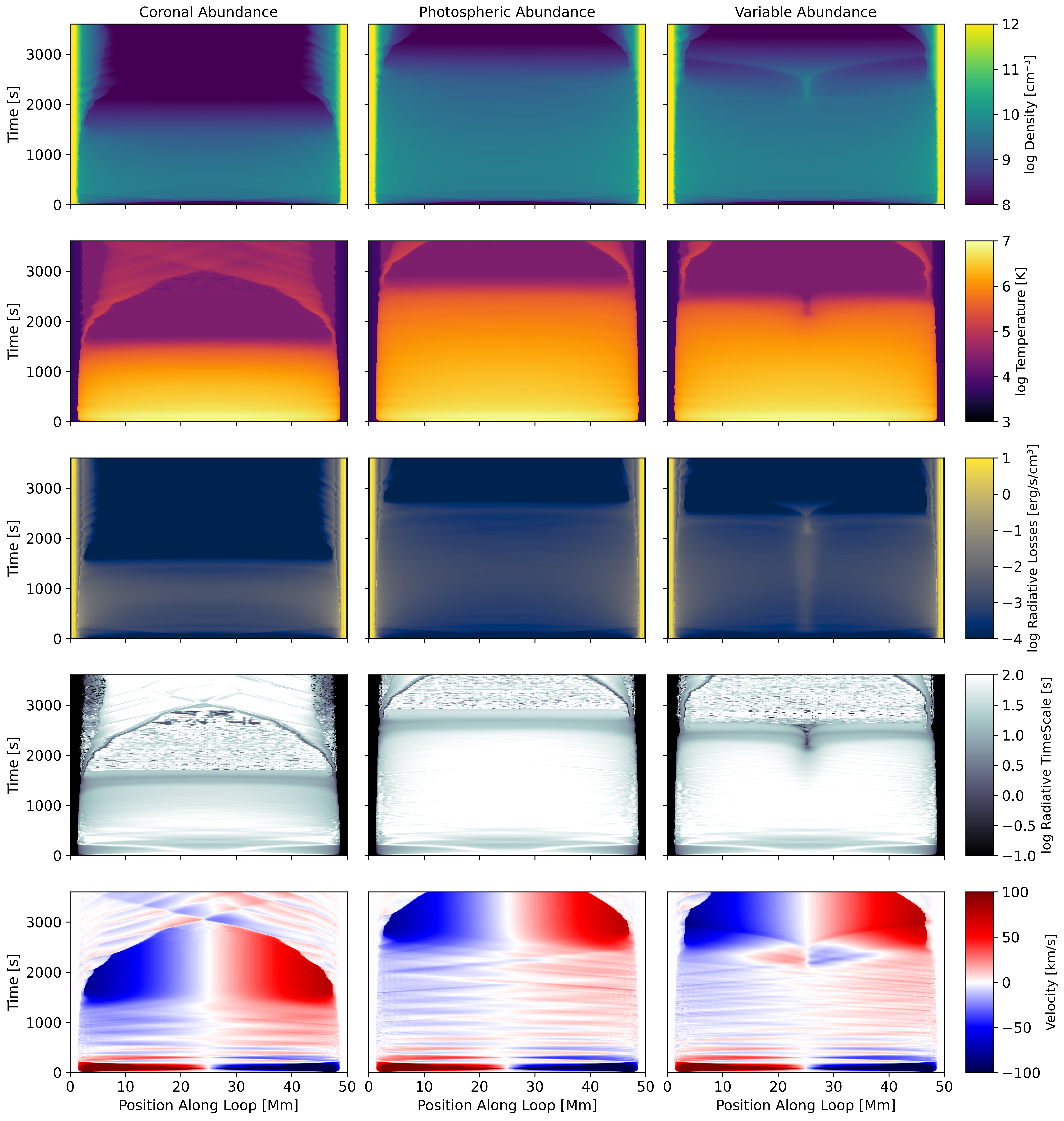}
    \caption{
    Impulsive nanoflare-heated coronal loop.  The rows of plots show electron density, electron temperature, radiative losses, radiative timescale, and bulk flow velocity along the loop length as a function of time. Positive velocity indicates flows moving towards the right and vice versa.
    The columns of plots show the three simulations, each with different assumptions of abundances (photospheric, coronal, and variable). The loop heats up and cools down almost uniformly, except in the model with variable elemental abundances which forms a condensation at the apex of the loop. 
    }
    \label{fig.impulsivePhys}
\end{figure*}

Figure~\ref{fig.impulsivePhys} compares the solutions for the three abundance models under the nanoflare heating scenario for electron density, electron temperature, radiative losses, radiative timescale (see the \hyperref[s.Appendix]{Appendix} for further discussion), and bulk flow velocity along the loop length as a function of time. The columns correspond to the abundances with which the simulations were run. For the cases with static coronal (left) or photospheric (center) abundances, the loop uniformly heats up and cools down which is typical of impulsively heated simulations \citep{Reep2020}. The photospheric case cools at a timescale almost double that of the coronal case. This is indicative of how radiation and radiative timescales can change simply due to the abundances although the dynamics remain comparable. Emissivity rapidly increases once temperature drops below $10^6$ K, and this rapid cooling across the whole loop, which also results in a rapid density decrease, causes the loop to catastrophically collapse. In all three cases, after the loop catastrophically collapses, the behavior of the system becomes non-physical due to numerical artifacts.

The inclusion of variable abundances (right) fundamentally alters the system's dynamics, and a coronal condensation forms around $t=2{,}100 $ seconds. 
The formation of coronal condensations proceeds  %similarity in both variable abundance simulations. 
as follows: compression from flows causes a local peak in $f$ that causes the radiation to increase locally. The spike in radiation causes the temperature to decrease at the apex faster than elsewhere in the corona, and the temperature decrease is equivalent to a pressure decrease ($P = 2 n k_B T$). The resulting pressure gradient force accelerates the plasma into the low pressure region, thus increasing the density. The increased density further increases the radiative loss rate ($R \propto n^2$; Equation~\ref{eq.BolLoss}), and, as radiation grows even stronger, the temperature decreases, and more material flows into the apex. The combined effects cause the cooling to run away, and the temperature drops incredibly fast, triggering a thermal runaway, which results in condensation.

%\newpage

\begin{figure*}[p]%[h!]
    \centering
    \includegraphics[width=1\textwidth]{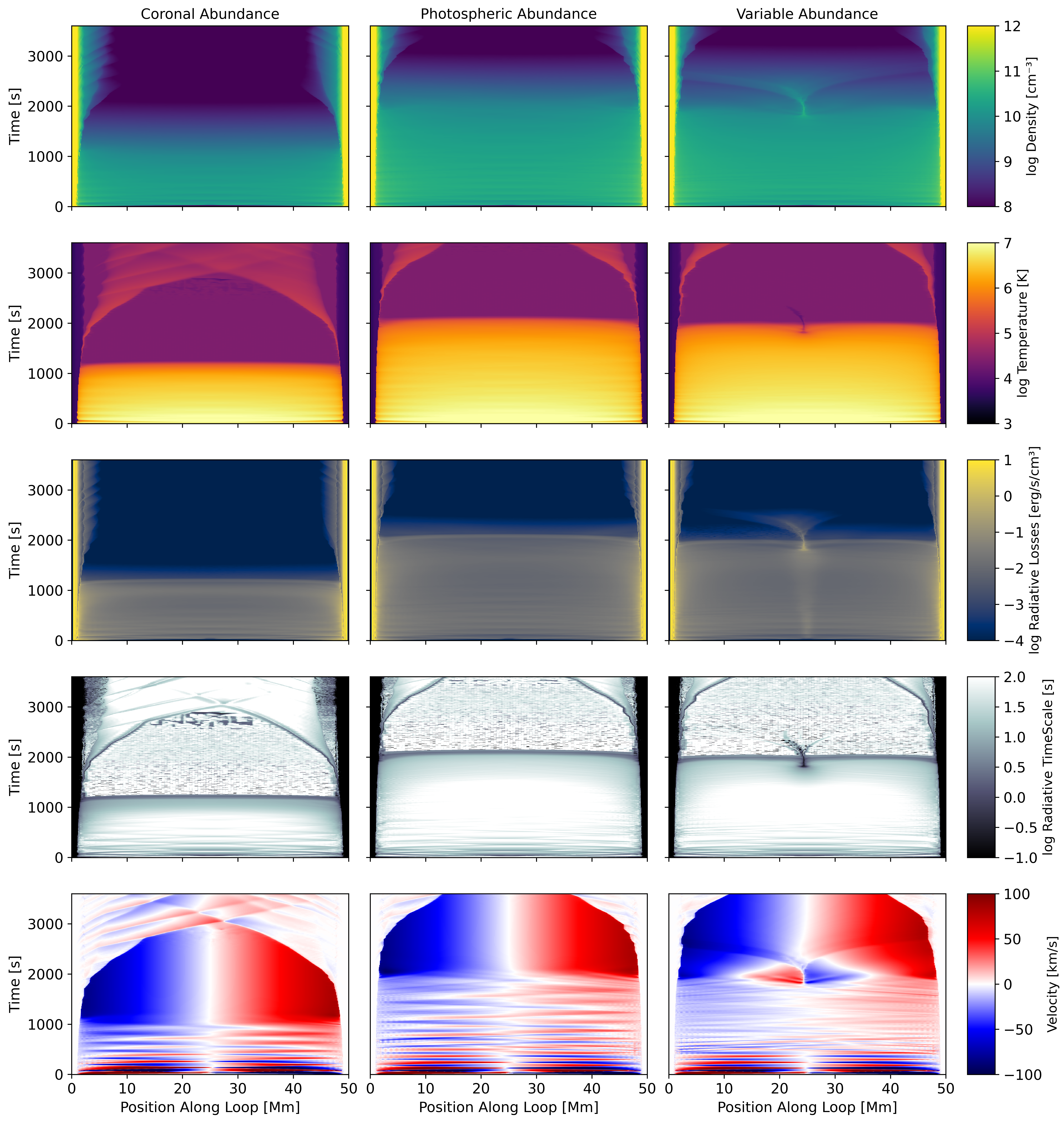}
    \caption{
    Electron beam-heated coronal loop, similar to Figure~\ref{fig.impulsivePhys}.  The loop heats up and cools down almost uniformly except in the HYDRAD model with variable elemental abundance which forms coronal rain at the apex of the loop.
    }
    \label{fig.beamPhys}
\end{figure*}

Similar to Figure~\ref{fig.impulsivePhys}, the results for the electron beam-heated loop are shown in Figure~\ref{fig.beamPhys}. Again, for the cases with fixed coronal or photospheric abundances, the loop uniformly heats up and cools. Cooling is initially due to conduction, then, as density increases and temperature decreases, radiation starts to dominate (see the \hyperref[s.Appendix]{Appendix} for further discussion). Much later in the simulation, the loop can cool mainly due to material draining from the loop \citep{bradshaw2010}. The different, but constant, abundance levels only affect the timescale of cooling. 
However, the variable abundance case, again, shows fundamentally different behavior, with coronal condensation forming around $t=1{,}800$ seconds. 
The fact that condensation manifests only for cases with variable abundance, regardless of the heating mechanism, is strong evidence that variable abundances are sufficient for the production of rain. 

The condensation in the beam heating case, in Figures \ref{fig.abdF} and \ref{fig.beamPhys}, has an asymmetric evolution. The simulations we present are set to be symmetric about the apex of the loop, but it is possible for asymmetries to arise. For example, at times steps where there are an odd number of grid cells due to AMR, the column density on each leg of the loop cannot be exactly the same; therefore the electron beam heating rate on either side of the loop cannot be exactly the same.

%\newpage
%\newpage

\begin{figure*}[p]%[h!]
    \centering
    \includegraphics[width=1\textwidth]{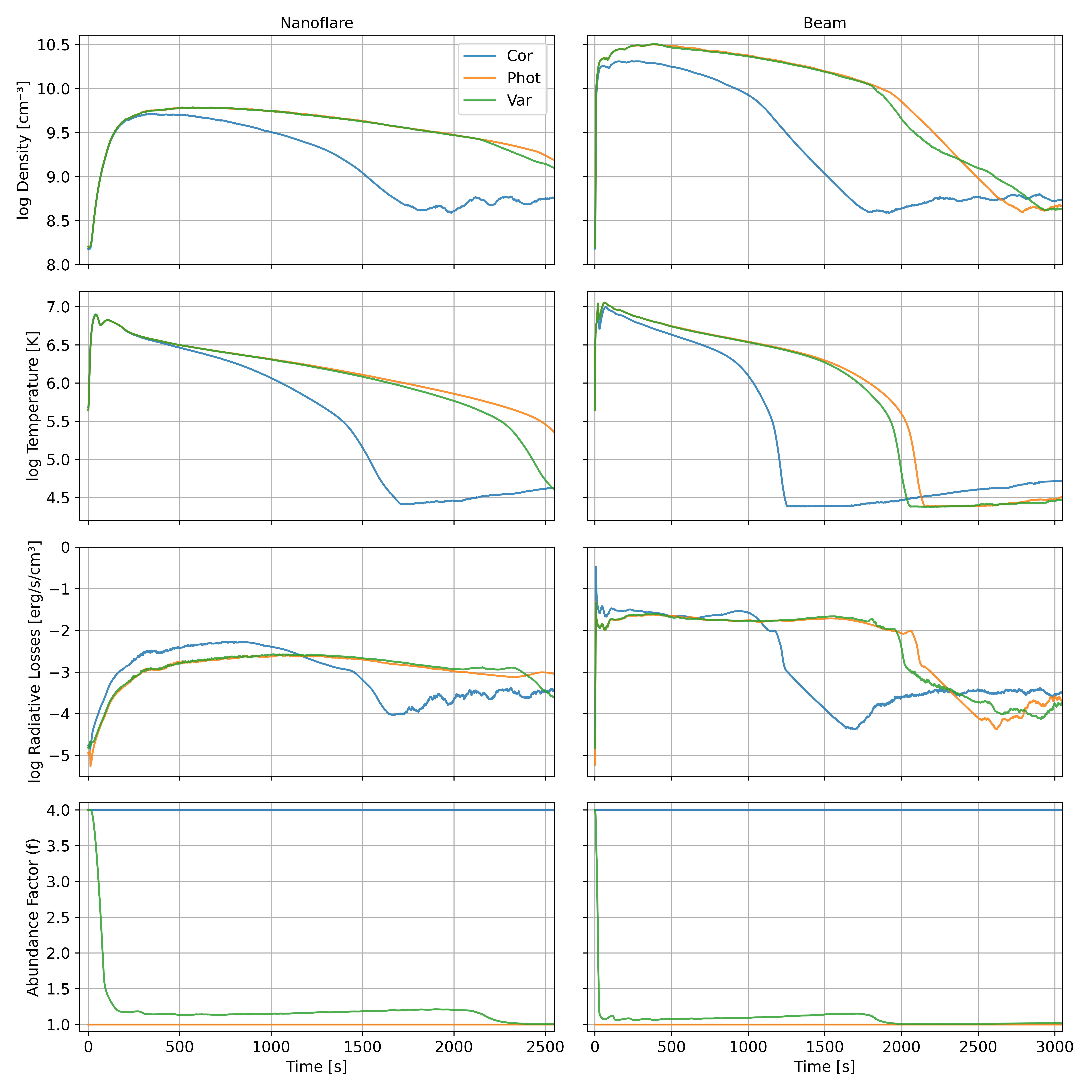}
    \caption{
    The rows of plots show the coronal averages of electron density, electron temperature, radiative losses, and abundance factor as a function of time. The left column shows the impulsive nanoflare-heated coronal loop, and the right column shows the electron beam-heated loop. 
    For the variable abundance case in the bottom row of plots, the initial drop in $f$ occurs as material flows into the loop, lowering its relative values. As outflows remove material, $f$ decreases further, resulting in a second, smaller, drop in the abundance factor associated with the condensation event.
    }
    \label{fig.ApexPlots}
\end{figure*}

We calculate the average values of electron density, electron temperature, bulk flow velocity, and abundance factor across the corona, $2.26<s<47.74$ Mm, in both sets of simulations. 
These coronal averages are shown in Figure \ref{fig.ApexPlots}. The rows show the different quantities as a function of time. The left column shows the impulsive nanoflare-heated coronal loop, and the right column shows the electron beam-heated loop. 
The initial drop in $f$ in the bottom row of plots occurs as material flows into the loop, reducing its relative value. When outflows remove material from the loop, $f$ continues to decrease, leading to a second, smaller, drop in the abundance factor associated with the condensation event. 
Similar to plots in \citet{Reep2024}, the curve denoting the simulations with variable abundance generally lies between the photospheric and coronal abundance cases. This is expected as the loop starts fractionated at coronal abundances ($f=4$) in the corona and decreases with time closer to photospheric abundances ($f=1$), so the overall behavior of the loop is between the two abundances.
%Recall that in all the abundance cases the behavior of the system is non-physical due to numerical artifacts once the loop catastrophically collapses.

\section{Discussion and Conclusion}
\label{s.Discussion and Conclusion}

Radiation plays a key role in plasma cooling, particularly in the solar corona. 
Observations reveal that elemental abundances fluctuate across different regions and over time \citep{DelZanna2021,Mondal2021,Nama2023,Mondal2023}, yet models typically treat them as constant over space and time \citep{Reep2020}. 
As a result, current impulsively heated models fail to replicate events such as rain that is present in both active regions and flares  \citep{Klimchuk2019,Reep2020,Antolin2020}. 
Since radiative energy loss is directly tied to these abundances, this simplification overlooks important physical processes in modeling the solar atmosphere. 

To improve how radiation is modeled, we have incorporated space and time-variable low FIP elemental abundances in the HYDRAD code. We have derived and implemented a continuity equation (Equation~\ref{eqf}) for low FIP elements, and we compare this new model with standard HYDRAD models that use abundances fixed in space and time. 

We have demonstrated that, by accounting for flows of low FIP elements, coronal condensations naturally form. In-flows from chromospheric evaporation lead to a reduction in $f$ along the coronal loop, except near its apex, where a relative peak in the abundance factor enhances radiative losses. The local enhancement in radiation at the apex of the loop directly causes condensations in impulsively heated loops.

At present, no other model treats radiation with spatiotemporally variable abundances. Spatiotemporal abundances are critical to understanding the cooling of plasma in the Sun's atmosphere and, as we have shown, can directly cause coronal rain. Implementing spatiotemporal abundances generalizes and improves the treatment of radiation in any magneto(hydrodynamic) model. 
Thus, investigating how $f(t,s)$ behaves when changing inputs such as heating rates and element sets needs to be explored further in the modified HYDRAD code as well as other magneto(hydrodynamic) models.
%codes like Bifrost \citep{Gudiksen2011}, LARE \citep{Arber2001}, and RADYN \citep{Carlsson1992,Carlsson1995,Carlsson1997,Allred2015}.

One method to validate the model would be comparing modeled events to EUV flare observations made with an imaging spectrometer, such as in \citet{To2024,Doschek2018}. They found that the loop-top brightens locally when observing low FIP elements' emissions, and that $f$ simultaneously increases at the apex of the loop. Although not shown explicitly in \citet{To2024}, they found no similar brightening in ions of high FIP elements such as Ar XIV (To 2024, private communication). Using the CHIANTI database, we can use outputs from the modified HYDRAD model and forward model synthesized emission for direct comparison. 

\citet{Brooks2024} observed coronal rain with Hinode/EIS, finding that rain has a complicated FIP bias, determined to be photospheric with a Si/S ratio, while the hotter temperature plasma around the rain was coronal as measured with a Ca/Ar ratio. This observation is consistent with the model presented in this paper. During chromospheric evaporation, the Si/S ratio becomes photospheric, while Ca/Ar remains coronal in the surrounding hot plasma. This suggests that the rain formation mechanism in \citet{Brooks2024} follows the same physics as our model, but further spectroscopic measurements could help to clarify this.

The HYDRAD code can be further improved by modeling the FIP effect. The simulations presented in this work assume that the coronal loop starts fractionated, but implementing a ponderomotive force will allow us to start at what would be $t=0$ before the heating event (\textit{i.e.}, magnetic reconnection creating Alfvén waves that result in a ponderomotive force). This would further generalize the modeling of abundances.

%\begin{acknowledgments}
\section*{Acknowledgments}
Support was provided by the International Space Science Institute (ISSI) in Bern, Switzerland, through ISSI International Team project \#545 ``Observe Local Think Global: What Solar Observations can Teach us about Multiphase Plasmas across Physical Scales," led by Drs. Clara Froment and Patrick Antolin.

We thank Craig Johnston, Cooper Downs, David Brooks, Clara Froment, Patrick Antolin, Steve Bradshaw, and the anonymous referee for their support through discussions regarding this work.

The HYDRAD simulations in this wo1rk were configured with the pydrad python package \citep{Barnes2023}, and the emissivity function in the appendix was calculated using the fiasco python package \citep{barnes2025}. The exact data used to produce the figures in the main text has been cached on Zenodo: doi:\href{https://doi.org/10.5281/zenodo.17103565}{10.5281/zenodo.17103565}
%\end{acknowledgments}

% \newpage

\appendix
\label{s.Appendix}
% \section{Appendix}
% \label{s.}

A coronal loop primarily cools through 3 processes: thermal conduction, radiation, and enthalpy via mass flows (evaporation/draining) \citep{Serio1991,Bradshaw2005,bradshaw2010,Cargill2013}. Initially, cooling is primarily done through conduction (high-$T$, high-$n$), and radiation (lower-$T$, high-$n$) becomes more important as the loop cools. As the loop cools even further, the loop drains its material proportionally to $L/v$.

The strength of a system's conduction can be assessed by examining its conduction timescale, \textit{i.e.}, the time it takes to lose the total energy of a given volume through conduction along magnetic field lines. For an ideal gas, this is given by

\begin{equation}
    \tau_C = 3k_B\kappa_0^{-1}T^{-5/2}nL^2
    \label{eq.condTime}
\end{equation}

\noindent where $k_B$ is the Boltzmann constant, $\kappa_0$ is the Spitzer thermal conductivity coefficient \citep{Spitzer1953}, $T$ is temperature, $n$ is the number density, and L is the half-length of the loop.

Similarly, the strength of a system's radiation can be assessed by examining its radiative timescale, \textit{i.e.}, the time it takes to radiate the total energy of a given volume. For example, \citet{Cargill1995} used a piecewise power-law fit using parameters $\alpha$ and $\chi$ for the emissivity function $\Lambda(T_{e}, n_{e}, f)$ in order to approximate radiative timescale. For an ideal gas, this is given by

\begin{equation}
    \tau_R = 3k_BT^{1-\alpha}n^{-1}\chi^{-1} %\frac{3k_BT^{1-\alpha}}{n \chi}
    \label{eq.radTime}
\end{equation}

\noindent where $\alpha$ and $\chi$ are dependent on the local elemental abundances (see \citealt{Klimchuk2008} as an example using coronal abundances). 

Emissivity $\Lambda(T_{e}, n_{e}, f)$ depends on elemental abundances, which is often overlooked and assumed to be constant (\textit{e.g.}, \citealt{Cargill1995} and \citealt{Klimchuk2008}; conversely, \citet{Brooks2018} uses 3 different static abundance cases). This discrepancy leads to systematic errors when calculating radiative losses, so accounting for varying elemental abundances is essential.

\begin{figure*}[h!]
    \centering
    \includegraphics[width=0.85\textwidth]{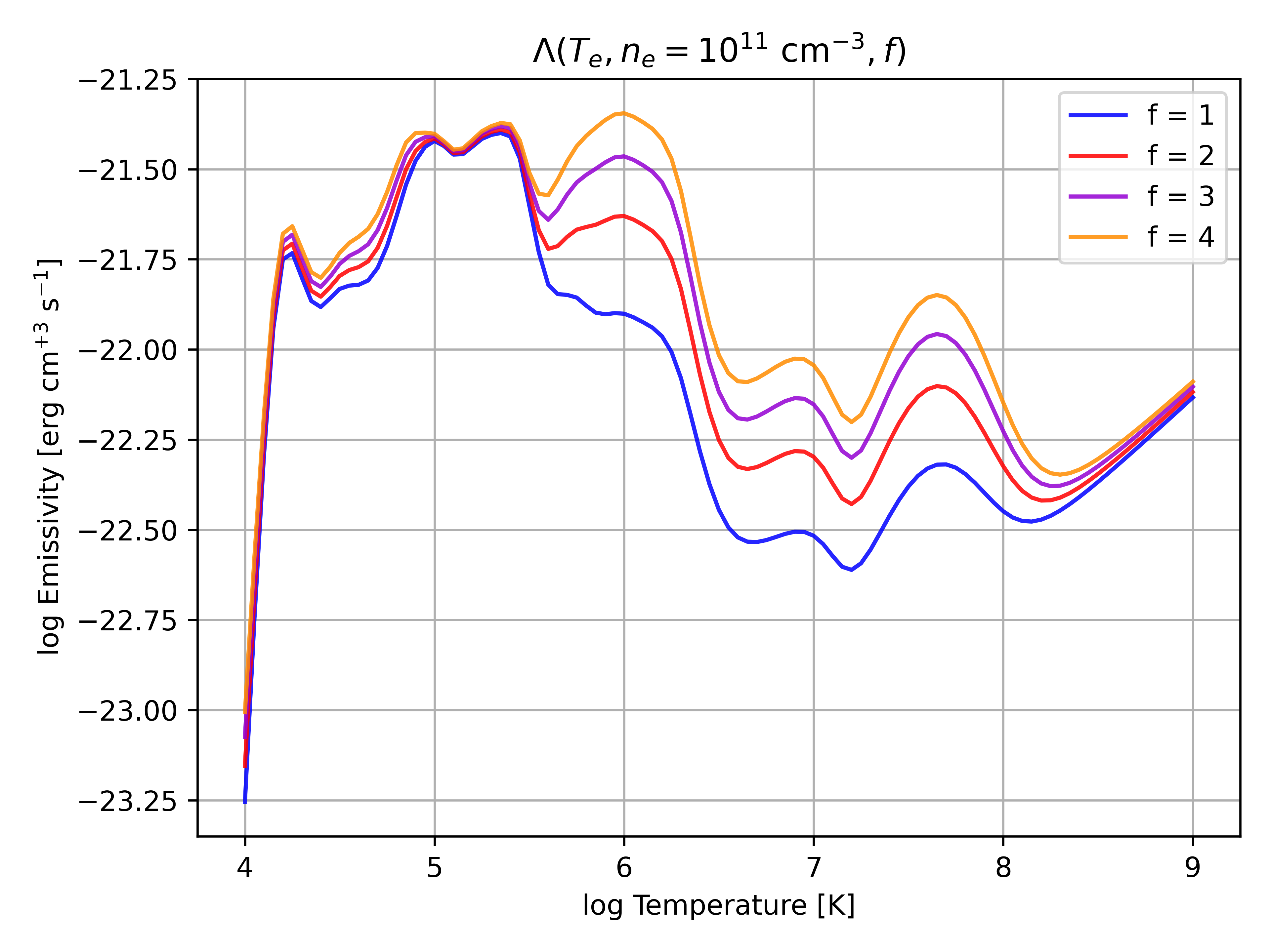}
    \caption{
    Emissivity $\Lambda(T_{e}, n_{e}, f)$ as a function of temperature for different abundance factors $f$, assuming $n_e = 10^{11}$ cm$^{-3}$.
    }
    \label{fig.loss_rate}
\end{figure*}

Accounting for varying elemental abundances, we fit a piecewise power-law to the emissivity function, shown in Figure~\ref{fig.loss_rate}, to find $\alpha$ and $\chi$ for different temperature ranges and $f$ values. 
We can then use these $\alpha$ and $\chi$ values, shown in Table~\ref{tab.timescales}, to approximate $\tau_R$ (\textit{e.g.}, \citealt{Cook1989}). As an example, we calculate $\tau_C$ and $\tau_R$, using values from Figure \ref{fig.ApexPlots}, for the beam heated loop assuming photospheric abundances:

\vspace{0.4cm}

\begin{itemize}
  \item $t = 300$s ($T\sim10^{6.8}$ K, $n\sim10^{10.5}$ cm$^{-3}$) : $\tau_C = 1{,}000$s and $\tau_R = 2{,}700$s
  \item $t = 500$s ($T\sim10^{6.7}$ K, $n\sim10^{10.5}$ cm$^{-3}$) : $\tau_C = 1{,}900$s and $\tau_R = 2{,}100$s
  \item $t = 1{,}000$s ($T\sim10^{6.5}$ K, $n\sim10^{10.4}$ cm$^{-3}$) : $\tau_C = 4{,}700$s and $\tau_R = 1{,}500$s
  \item $t = 1{,}500$s ($T\sim10^{6.3}$ K, $n\sim10^{10.2}$ cm$^{-3}$) : $\tau_C = 9{,}300$s and $\tau_R = 650$s
\end{itemize}

We also calculate $\tau_R$ as a function of temperature. For example, the coronal condensations presented in Figures \ref{fig.impulsivePhys} and \ref{fig.beamPhys} have an electron density of $n_e\sim10^{11}$ cm$^{-3}$ and electron temperature of $T_e\sim3.5\cdot10^{4}$ K. $\tau_R$, assuming the density of a coronal condensation, is as follows:

\vspace{0.4cm}

\begin{itemize}
  \item $T_e = 10^{7}$ K : $\tau_R(f=1) = 1{,}400$s and $\tau_R(f=4) = 420$s
  \item $T_e = 10^{6}$ K : $\tau_R(f=1) = 32$s and $\tau_R(f=4) = 8.8$s
  \item $T_e = 10^{5}$ K : $\tau_R(f=1) = 1.3$s and $\tau_R(f=4) = 1.2$s
  \item $T_e = 3.5\cdot10^{4}$ K : $\tau_R(f=1) = 0.76$s and $\tau_R(f=4) = 0.66$s
\end{itemize}

\noindent So, as the plasma associated with the coronal condensation becomes cooler, the radiative timescale becomes smaller, and, in extreme cases, can be smaller than $1$s. HYDRAD directly calculates $\tau_R$, as seen in Figures \ref{fig.impulsivePhys} and \ref{fig.beamPhys}, without approximating $\Lambda(T_{e}, n_{e}, f)$.

\begin{table}%[!htbp]
  \centering
    \begin{tabular}{l|cccccccc}
    \toprule
    %\hline
    Temperature Range & $\alpha(f = 1)$ & $\chi(f = 1)$ & $\alpha(f = 2)$ & $\chi(f = 2)$ & $\alpha(f = 3)$ & $\chi(f = 3)$ & $\alpha(f = 4)$ & $\chi(f = 4)$ \\
    %\midrule
    \hline
    4.00 $\leq \log T \leq$ 4.16 & 8.06 & 3.56e-56 & 7.78 & 6.05e-55 & 7.57 & 4.94e-54 & 7.41 & 2.51e-53 \\
    4.16 $\leq \log T \leq$ 5.37 & 0.46 & 1.54e-24 & 0.43 & 2.20e-24 & 0.41 & 3.06e-24 & 0.38 & 4.14e-24 \\
    5.37 $\leq \log T \leq$ 5.56 & -2.30 & 9.69e-10 & -1.96 & 1.49e-11 & -1.65 & 3.40e-13 & -1.38 & 1.29e-14 \\
    5.56 $\leq \log T \leq$ 5.93 & -0.36 & 1.58e-20 & 0.21 & 1.25e-23 & 0.51 & 3.22e-25 & 0.70 & 3.34e-26 \\
    5.93 $\leq \log T \leq$ 6.23 & -0.13 & 7.59e-22 & -0.20 & 3.82e-21 & -0.25 & 1.08e-20 & -0.28 & 2.25e-20 \\
    6.23 $\leq \log T \leq$ 6.53 & -1.88 & 5.60e-11 & -2.12 & 3.10e-09 & -2.21 & 1.78e-08 & -2.26 & 4.85e-08 \\
    6.53 $\leq \log T \leq$ 6.98 & 0.01 & 2.65e-23 & 0.10 & 1.06e-23 & 0.14 & 7.82e-24 & 0.16 & 7.02e-24 \\
    6.98 $\leq \log T \leq$ 7.19 & -0.49 & 7.98e-20 & -0.73 & 6.75e-18 & -0.85 & 6.50e-17 & -0.92 & 2.72e-16 \\
    7.19 $\leq \log T \leq$ 7.67 & 0.65 & 4.98e-28 & 0.77 & 1.12e-28 & 0.83 & 5.02e-29 & 0.88 & 2.98e-29 \\
    7.67 $\leq \log T \leq$ 8.24 & -0.35 & 2.30e-20 & -0.69 & 1.76e-17 & -0.90 & 1.01e-15 & -1.05 & 1.67e-14 \\
    8.24 $\leq \log T \leq$ 9.00 & 0.48 & 3.50e-27 & 0.44 & 7.91e-27 & 0.40 & 1.84e-26 & 0.37 & 4.06e-26 \\ 
    \bottomrule
    %\hline
    \end{tabular}
  \caption{For a given temperature range and abundance factor, the fitted values of $\alpha$ and $\chi$ are shown, assuming $n_e = 10^{11}$ cm$^{-3}$.  The top row shows different abundance factors, and the left column shows different temperature ranges. }
  \label{tab.timescales}
\end{table}

\newpage

\bibliography{tVariableAbundances}{}
\bibliographystyle{aasjournal}

\end{document}